\documentclass[journal]{IEEEtran}
\usepackage{amsmath,amsfonts}
\usepackage{graphicx}
\usepackage{cite}
\usepackage{booktabs}
\usepackage{url}
\usepackage{stfloats}
\usepackage[caption=false,font=footnotesize]{subfig}
\usepackage{algorithm}
\usepackage{algorithmic}
\usepackage{tabularx}
\usepackage{array}
\usepackage{makecell}
\begin{document}

\title{High-Fidelity Digital Twin Dataset Generation for Inverter-Based Microgrids Under Multi-Scenario Disturbances}

\author{Osasumwen Cedric Ogiesoba-Eguakun,~\IEEEmembership{Member,~IEEE,}, Kaveh Ashenayi, ~\IEEEmembership{Senior Member,~IEEE}, \\ Suman Rath,~\IEEEmembership{Member,~IEEE}
\thanks{This work was conducted as part of the graduate research activities at the University of Tulsa.\\ 
(Corresponding author: Osasumwen Cedric Ogiesoba-Eguakun.)
 
O. C. Ogiesoba-Eguakun, K. Ashenayi, and S. Rath are with the Department of Electrical and Computer Engineering, The University of Tulsa, Tulsa, OK 74104, USA (e-mail: oco1411@utulsa.edu, kash@utulsa.edu, suman-rath@utulsa.edu).}

}



\maketitle

\begin{abstract}
Public power-system datasets often lack electromagnetic transient (EMT) waveforms, inverter control dynamics, and diverse disturbance coverage, which limits their usefulness for training surrogate models and studying cyber-physical behavior in inverter-based microgrids. This paper presents a high-fidelity digital twin dataset generated from a MATLAB/Simulink EMT model of a low-voltage AC microgrid with ten inverter-based distributed generators. The dataset records synchronized three-phase PCC voltages and currents, per-DG active power, reactive power, and frequency, together with embedded scenario labels, producing 38 aligned channels sampled at $\Delta t = 2~\mu$s over $T = 1$~s ($N = 500{,}001$ samples) per scenario. Eleven operating and disturbance scenarios are included: normal operation, load step, voltage sag (temporary three-phase fault), load ramp, frequency ramp, DG trip, tie-line trip, reactive power step, single-line-to-ground faults, measurement noise injection, and communication delay. To ensure numerical stability without altering sequence length, invalid samples (NaN, Inf, and extreme outliers) are repaired using linear interpolation. Each scenario is further validated using system-level evidence from mean frequency, PCC voltage magnitude, total active power, voltage unbalance, and zero-sequence current to confirm physical observability and correct timing. The resulting dataset provides a consistent, labeled EMT benchmark for surrogate modeling, disturbance classification, robustness testing under noise and delay, and cyber-physical resilience analysis in inverter-dominated microgrids. The dataset and processing scripts will be released upon acceptance. 
\end{abstract}


\section{Introduction}
\IEEEPARstart{M}{odern} microgrids are becoming increasingly inverter-dominated due to the high use of photovoltaic (PV) systems, battery energy storage, and converter-interfaced resources \cite{ogiesoba2023design,al2025digital}. These systems can rapidly change operating modes (grid-connected, islanded, and resynchronized), and their behavior is strongly influenced by inverter control loops, protection actions, and fast electromagnetic transients. Because of this, many studies on stability, protection, and cyber-physical resilience now require EMT level signals rather than slow phasor or steady-state measurements. As inverter-based resources continue to grow, microgrid stability and protection increasingly depend on fast control interactions and electromagnetic transient behavior, which cannot be captured by traditional steady-state tools such as MATPOWER \cite{zimmerman2010matpower}.

At the same time, most publicly available power-system datasets are not built for these needs. Common test-case libraries are very useful for power-flow and planning studies, but they do not include microsecond-level dynamics, inverter inner-loop behavior, or realistic transient waveforms needed for EMT benchmarking \cite{lin2020research}. Even when time-series datasets exist, they are often sampled slowly and provide only system-level measurements, which limits their usefulness for studying fast inverter-dominated microgrid behavior.

This limitation is important because data-driven surrogate modeling is now widely used for fast prediction and real-time decision support in power systems. Surrogate models can greatly reduce simulation time, but their accuracy depends strongly on the quality of the training data, including disturbance diversity, consistent labeling, and physical realism. Recent surveys on surrogate modeling show that reliable models depend on carefully generated datasets that capture real system physics and operating conditions \cite{aghazadeh2024digital, jafari2023review}. At the same time, digital twin frameworks are now widely used to connect detailed simulations with real-time monitoring and control. However, they depend on realistic datasets that include disturbances and clearly measurable system responses \cite{mohammadi2024surrogate}.

To address these gaps, this work presents a high-fidelity digital-twin dataset generated from a detailed inverter-based microgrid model, designed specifically for surrogate modeling and cyber-physical analysis \cite{ogiesoba2026cyberattack, cedric2025quantum}. The dataset is created using EMT-style simulation and provides synchronized multi-channel measurements, including three-phase voltage and current at the point of common coupling (PCC), as well as per-distributed-generation active power, reactive power, and frequency. Eleven labeled operating and disturbance scenarios are included, covering common power events (load steps, ramps, voltage sag and fault events, generator trips) together with data-quality and cyber-physical stressors such as noise injection and communication delay.

A key design goal is that each scenario is not only labeled but also validated using system-level evidence, including frequency trajectories, PCC voltage behavior, and total active power response. This follows practices in real-time simulation and hardware-in-the-loop studies, where signal timing, integrity, and closed-loop response must be demonstrated rather than assumed \cite{thwe2025digital, shen2024virtual}. Prior reviews of HIL and digital-twin applications further highlight the need for realistic datasets when controllers, protection devices, phasor measurement units (PMUs), and communication effects are involved \cite{von2023power}.

The main contributions of this paper are:
\begin{itemize}
    \item A labeled, multi-scenario EMT microgrid dataset designed for surrogate modeling and cyber-physical benchmarking.
    \item A consistent dataset structure with synchronized measurement channels across all scenarios to support learning and evaluation.
    \item Scenario-by-scenario validation evidence, including plots and summary statistics, to confirm that each disturbance is physically observable in the exported data.
\end{itemize}

The remainder of this paper describes the microgrid model and measurement channels, the scenario design and labeling rules, the data export and cleaning workflow, the validation evidence for each scenario, and how the dataset supports the next phase of training and evaluating surrogate models for fast microgrid prediction. 

\section{Related Work}

Research on power-system data and models covers many applications, ranging from planning studies to dynamic simulation and machine-learning integration. Public datasets such as the Open Power System Data time-series collection provide aggregated load, generation, and market data for planning studies. However, these datasets do not include high-frequency EMT waveforms, which are required for studying inverter-dominated microgrids and training dynamic surrogate models \cite{OPSD2026,aghdam2025navigating}.

Digital twin technologies have emerged as an important approach for representing physical energy systems in virtual environments. Digital twins support monitoring, control, and operational decision-making by combining sensor data, simulation models, and analytics. Recent reviews show that digital twins are increasingly used in smart energy systems to improve monitoring, prediction, and system resilience \cite{aghazadeh2024digital, mbasso2025digital}. Several studies also show how digital twins are used in microgrids for energy management and real-time simulation of distributed energy resources (DERs) \cite{shao2023digital,jiang2024digital}. These works confirm that detailed models can reproduce system behavior. However, most digital twin studies focus on framework development or control applications and do not provide labeled, synchronized, high-rate waveform datasets for machine-learning tasks.

Surrogate modeling and reduced-order dynamic modeling have also gained attention as tools for reducing computational cost while preserving key system dynamics. Prior studies emphasize that data-driven surrogate models depend strongly on dataset quality, time resolution, and disturbance diversity. When training data lack rich transient events, surrogate models may perform poorly during faults, inverter control interactions, or communication disturbances \cite{nuevo2025real}.

Overall, existing studies show three major gaps:

\begin{itemize}
    \item Most public datasets do not provide high-resolution EMT waveforms for multiple disturbance scenarios \cite{sticht2024power}.
    \item Digital twin studies often stop at architecture or application without releasing benchmark datasets for machine learning \cite{barreto2025cyber, aravena2025open}.
    \item Surrogate model development is limited by the lack of validated, multi-scenario dynamic data \cite{oelhaf2025scoping}.
\end{itemize}

The dataset presented in this paper addresses these limitations by providing synchronized, labeled, multi-scenario waveform measurements designed specifically for surrogate modeling and cyber-physical benchmarking.

\begin{table}[!t]
\caption{Comparison of Representative Power-System Datasets and Digital Twin Studies}
\label{tab:relatedwork}
\centering
\resizebox{\columnwidth}{!}{%
\begin{tabular}{lcccc}
\hline
Ref. & System Type & Data Resolution & Disturbances & Public Data \\
\hline
\cite{zimmerman2010matpower} & Transmission & Static & None & Yes \\
\cite{OPSD2026} & Bulk grid & Minutes--hours & Limited & Yes \\
\cite{shao2023digital} & Microgrid & Seconds--ms & Load changes & No \\
\cite{jiang2024digital} & Microgrid & ms & Control events & No \\
\cite{aghazadeh2024digital} & Review & Mixed & Not specified & No \\
\cite{nuevo2025real} & Review & ms & Limited dynamics & No \\
This work & Inverter microgrid & \textbf{Microseconds} & \textbf{11 scenarios} & \textbf{Yes} \\
\hline
\end{tabular}%
}
\end{table}

Table~\ref{tab:relatedwork} compares representative public datasets and digital twin studies. Most existing datasets focus on steady-state or slow dynamics and offer limited disturbance diversity. Prior digital twin studies mainly emphasize system modeling and control, but do not release labeled EMT waveform datasets. In contrast, this work provides synchronized EMT measurements with multiple disturbance scenarios, designed specifically for surrogate modeling and cyber-physical benchmarking.

\section{Digital Twin Microgrid Model and Measurement Architecture}

This section explains the inverter-based microgrid digital twin used to generate the dataset. It presents the system configuration, distributed generation structure, control design, and measurement channels.

\subsection{Microgrid Configuration}

The digital twin models a low-voltage alternating current (AC) microgrid made up of ten inverter-based distributed generation (DG) units. Each unit represents a renewable or storage resource connected through a power converter. A PCC links the microgrid to the upstream utility grid through a controllable tie line, allowing both grid-connected and islanded operation.

Each inverter includes inner current and voltage control loops, outer power-frequency and reactive-voltage droop control, and phase-locked loop synchronization \cite{rajendran2025brain, pogaku2007modeling}. This structure follows conventional grid-following inverter architectures widely adopted in practical microgrids. Such detailed control modeling is required to capture fast EMT behavior and inverter interactions that cannot be represented using phasor or averaged models.

The microgrid also includes aggregated static and dynamic loads distributed across the AC bus. Protection logic and breaker models are included to allow generator trips, tie-line disconnection, and fault scenarios. This configuration allows realistic operating transitions, including load changes, voltage sags, frequency deviations, and islanding events.

The full system is implemented in MATLAB/Simulink using EMT-style simulation with a fixed time step of $\Delta t = 2~\mu$s. All logged signals are sampled at the EMT step without down-sampling, producing discrete-time sequences
\begin{equation}
x[n] = x(n\Delta t), \quad n = 0,1,\ldots,N-1,
\label{eq:emt_sampling}
\end{equation}
which preserve fast voltage, current, and control-loop dynamics. This follows established practice in real-time and hardware-in-the-loop research \cite{von2023power, thwe2025digital, palensky2017cosimulation}.

\begin{figure*}[!t]
    \centering
    \includegraphics[width=\textwidth]{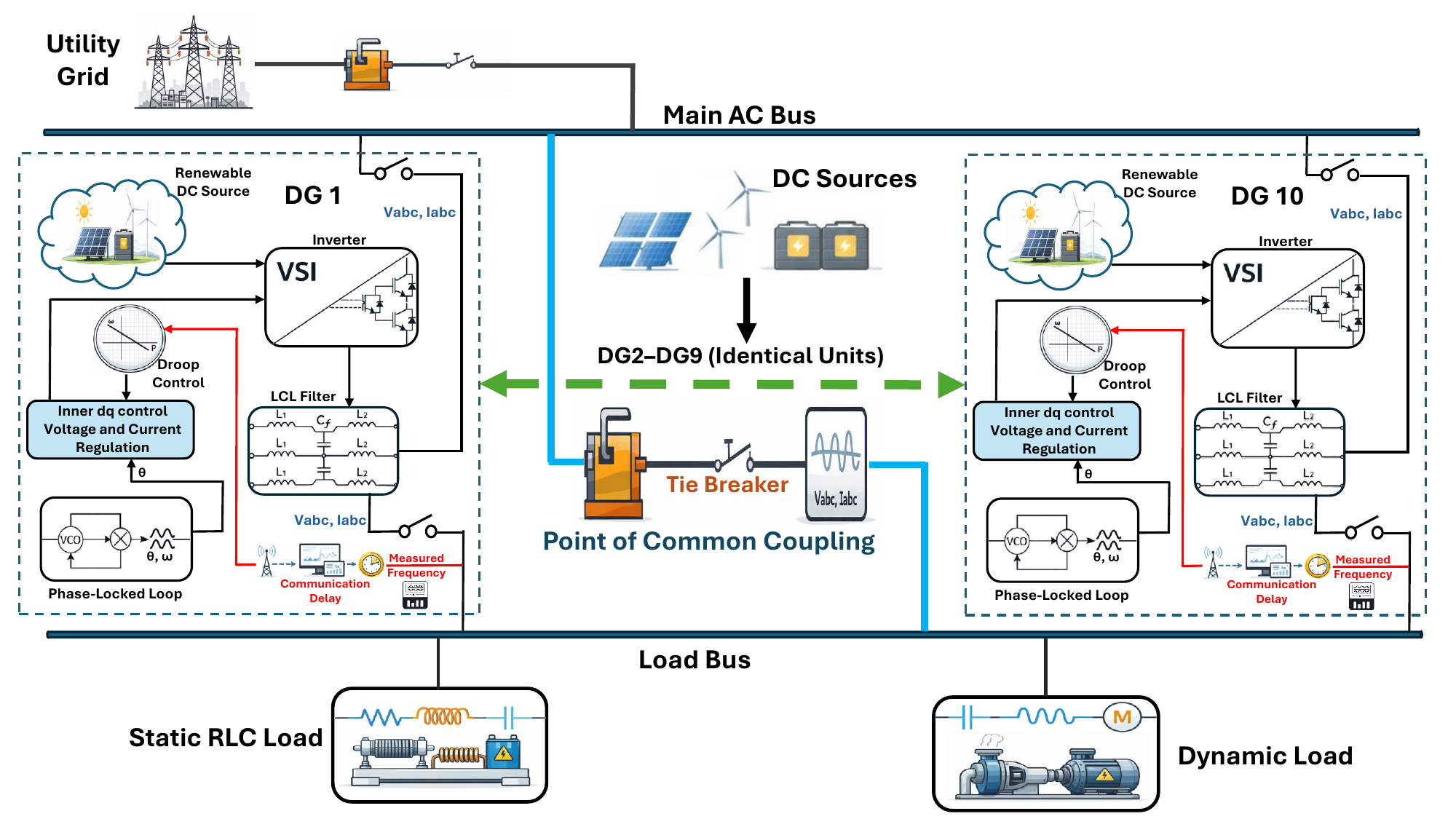}
    \caption{Single-line diagram of the inverter-based microgrid digital twin used for dataset generation. The utility grid is connected at the PCC through a tie-line breaker to enable grid-connected and islanded operation. Ten identical inverter-based DG units (DG1--DG10) connect to the main AC bus, supplying static resistor–inductor–capacitor (RLC) and dynamic loads. Logged measurements include PCC three-phase voltage and current and per-DG active power, reactive power, and frequency.}
    \label{fig:microgrid_sld}
\end{figure*}

Figure~\ref{fig:microgrid_sld} shows the digital-twin microgrid topology, PCC interface, DG connections, and load placement used in all dataset scenarios.

\subsection{Distributed Generation and Control Structure}

Each DG unit supplies active and reactive power to the AC microgrid through a voltage source inverter (VSI). The inverter uses two control levels: an outer droop-based power controller and inner current control loops in the synchronous reference frame.

The outer control layer sets the frequency and voltage using standard droop rules. Active power is controlled through frequency droop, while reactive power is controlled through voltage droop, as expressed in (\ref{eq:droop_pf})–(\ref{eq:droop_qv}).

The inner control loops follow these references by controlling the inverter output currents. This allows the inverter to respond quickly to changes and keep the voltage at its terminals stable.

Using this control structure, DG units share load automatically without communication while maintaining stable system frequency and voltage under varying operating conditions \cite{guerrero2010hierarchical}. Power sharing is governed by standard droop laws:

\begin{equation}
f_{DGk}(t) = f_0 - m_p\big(P_{DGk}(t) - P_{DGk}^{\ast}\big),
\label{eq:droop_pf}
\end{equation}

\begin{equation}
V_{DGk}(t) = V_0 - m_q\big(Q_{DGk}(t) - Q_{DGk}^{\ast}\big),
\label{eq:droop_qv}
\end{equation}

where $f_0$ and $V_0$ are the nominal frequency and voltage, $m_p$ and $m_q$ are the droop coefficients, and $P_{DGk}^{\ast}$ and $Q_{DGk}^{\ast}$ are the active and reactive power reference setpoints for DG $k$.

All DG controllers use the same parameter settings to keep their behavior consistent, while still allowing natural system interactions during disturbances. This makes it possible to study how the inverters respond together during events such as frequency ramps, reactive power changes, and communication delays.

Unlike simplified average inverter models commonly used in planning studies, the adopted EMT inverter models explicitly capture control-loop dynamics and transient behavior, which are essential for surrogate model training and cyber-physical benchmarking \cite{castilla2018control}.

\subsection{Measurement Channels and Dataset Signals}

Synchronized measurements are recorded across the microgrid to form the dataset. At the PCC, three-phase voltages $(V_1,V_2,V_3)$ and currents $(I_1,I_2,I_3)$ represent system-level electrical behavior \cite{terzija2010wide}. For each DG unit, active power $P_{DGk}$, reactive power $Q_{DGk}$, and frequency $f_{DGk}$ are recorded, where $k=1,\ldots,10$. When the electrical angular frequency is available from the inverter control model, the frequency is obtained as
\begin{equation}
f_{DGk}(t) = \frac{\omega_{DGk}(t)}{2\pi}.
\label{eq:freq_from_omega}
\end{equation}

Furthermore, the total system active and reactive power are obtained by summing the contributions of all DG units:
\begin{equation}
P_{\text{total}}(t) = \sum_{k=1}^{10} P_{DGk}(t), \qquad
Q_{\text{total}}(t) = \sum_{k=1}^{10} Q_{DGk}(t).
\label{eq:ptotal_qtotal}
\end{equation}
and the mean microgrid frequency is
\begin{equation}
f_{\text{mean}}(t) = \frac{1}{10} \sum_{k=1}^{10} f_{DGk}(t).
\label{eq:fmean}
\end{equation}

To check voltage behavior and identify stress at the PCC, a voltage-magnitude proxy is calculated using the three-phase voltage measurements:
\begin{equation}
V_{\text{PCC}}(t)=\sqrt{\frac{V_1^2(t)+V_2^2(t)+V_3^2(t)}{3}}.
\label{eq:vpcc_proxy}
\end{equation}

All signals are exported at the EMT simulation time step without down-sampling, resulting in microsecond-resolution time series. The dataset, therefore, includes time stamps, PCC measurements, per-DG electrical quantities, and scenario labels in a unified structure.

Such synchronized, high-rate multi-channel measurements are rarely available in public datasets, yet they are essential for learning fast inverter dynamics and transient responses. Prior digital twin studies mainly focus on model development or control applications and do not release labeled EMT waveform datasets for machine-learning tasks \cite{aghazadeh2024digital, shao2023digital}.

\begin{table}[!t]
\caption{Measurement channels and dataset schema (38 synchronized channels).}
\label{tab:schema_channels}
\centering
\renewcommand{\arraystretch}{1.25} 
\resizebox{\columnwidth}{!}{%
\begin{tabular}{llcc}
\toprule
\textbf{Group} & \textbf{Channels} & \textbf{Count} & \textbf{Units / Notes} \\
\midrule
Time & $t$ & 1 & s, sampled at $\Delta t = 2~\mu$s \\

PCC voltage & $(V_1,V_2,V_3)$ & 3 & \makecell[l]{V (three-phase, PCC)} \\

PCC current & $(I_1,I_2,I_3)$ & 3 & \makecell[l]{A (three-phase, PCC)} \\

DG active power & $(P_{DG1}\ldots P_{DG10})$ & 10 & \makecell[l]{W or kW (per-DG)} \\

DG reactive power & $(Q_{DG1}\ldots Q_{DG10})$ & 10 & \makecell[l]{var or kvar (per-DG)} \\

DG frequency & $(f_{DG1}\ldots f_{DG10})$ & 10 & \makecell[l]{Hz (per-DG)} \\

Scenario label & $y$ & 1 & \makecell[l]{integer $y\in\{0,\ldots,10\}$} \\

\midrule
\textbf{Total} &  & \textbf{38} & \makecell[l]{$N=500{,}001$ samples \\ per scenario ($T=1$ s)} \\
\bottomrule
\end{tabular}%
}
\end{table}

Table~\ref{tab:schema_channels} shows how the dataset is organized, including the different measurement groups, channel names, units, and total number of signals. Each scenario includes 38 synchronized signals sampled at a fixed time step, which keeps the data consistent and easy to reproduce across all disturbance cases.

\subsection{Scenario Labeling and Timing Control}

Each simulation run corresponds to a single operating or disturbance scenario. A global label signal is generated inside the digital twin and exported together with electrical measurements, providing precise timing of scenario onset and duration. Labels are synchronized with physical events such as load steps, fault insertion, generator trips, and communication delays.

This built-in labeling lets machine-learning models directly link signal patterns to known system conditions, without relying on post-processing or heuristic labeling.

\section{Scenario Design and Labeling Strategy}

The dataset includes eleven operating and disturbance scenarios that capture both fast electrical transients and control effects in an inverter-based microgrid. The scenarios include electrical disturbances (faults, load changes, and voltage or frequency variations), operating changes (DG trips and tie-line disconnection), and data-related effects (measurement noise and communication delay). Surrogate models perform better when trained on realistic operating behavior and disturbance responses, making this range of conditions important for data-driven modeling \cite{aghazadeh2024digital}. Each scenario is simulated independently and exported as a synchronized, fixed-length time series segment with an embedded label.

\subsection{Scenario Definitions}

\subsubsection{Normal Operation}
Normal grid-connected operation is simulated with steady loading and droop-controlled DG units. Voltage and frequency stay close to their nominal values, providing a baseline reference for all disturbed cases.

\subsubsection{Load Step}
A sudden increase in load is applied at the AC bus. This causes inverter currents and DG power to increase immediately. The system frequency drops for a short time and then returns to normal as droop control reacts.

\subsubsection{Voltage Sag}
A balanced three-phase fault is applied at the PCC to briefly reduce the voltage. The fault is then cleared, allowing us to see how the inverters respond and recover.

\subsubsection{Load Ramp}
The load is increased slowly over a set time period, causing a smooth rise in total active power demand. This scenario tests how well the system tracks gradual changes and handles slow transients.

\subsubsection{Frequency Ramp}
A controlled frequency variation is introduced through the grid reference, producing a monotonic drift in the measured DG frequencies. This scenario stresses synchronization behavior and droop response under abnormal frequency conditions.

\subsubsection{Generator (DG) Trip}
One DG unit is disconnected during operation. Its active and reactive power drops suddenly, while the remaining DG units share the load through droop control, causing a brief change in system frequency.

\subsubsection{Tie-Line Trip}
The tie line connecting the microgrid to the utility grid is opened, forcing a transition from grid-connected to islanded operation. This produces a distinct change in frequency regulation responsibility and power balance.

\subsubsection{Reactive Power Event}
A reactive power step is introduced by modifying reactive demand or a reactive command, producing a change in total reactive power and a voltage response at the PCC. This scenario emphasizes voltage regulation and reactive power sharing.

\subsubsection{Single-Line-to-Ground Fault}
Single-line-to-ground faults are applied on phases A, B, and C in separate runs. These faults create unbalanced voltages and currents and add zero-sequence components that can be seen in the PCC measurements.

\subsubsection{Noise Injection}
Additive measurement noise is injected into selected signals to represent sensor noise and imperfect acquisition. This scenario supports robustness tests under degraded measurement quality.

\subsubsection{Communication Delay}
A fixed delay is inserted in the frequency feedback path to represent communication latency between controllers. The resulting effect appears as subtle deviations in frequency and power-sharing dynamics rather than sharp electrical transients, consistent with digital twin and HIL observations \cite{von2023power, palensky2017cosimulation}.

\begin{table}[!t]
\caption{Scenario label mapping and deterministic event timing used for dataset generation.}
\label{tab:scenario_labels}
\centering
\renewcommand{\arraystretch}{1.25}
\resizebox{\columnwidth}{!}{%
\begin{tabular}{c l l p{5.2cm}}
\toprule
\textbf{Label} & \textbf{Scenario} & \textbf{Event time / window} & \textbf{Primary observable evidence} \\
\midrule
0  & Normal operation & none & stable $f_{\text{mean}}(t)$, $V_{\text{PCC}}(t)$, $P_{\text{total}}(t)$ \\

1  & Load step & $\approx 0.70$ s & step in $P_{\text{total}}(t)$, dip in $f_{\text{mean}}(t)$, rise in PCC current \\

2  & \makecell[l]{Voltage sag \\ (3$\phi$ fault at PCC)} & sag window & drop in $V_{\text{PCC}}(t)$ with aligned change in $P_{\text{total}}(t)$ \\

3  & Load ramp & 0.50--0.70 s & monotonic increase in $P_{\text{total}}(t)$ with small coordinated frequency deviation \\

4  & Frequency ramp & 0.50--0.70 s & ramp in $f_{DGk}(t)$ with distinct $df/dt$ signature during the window \\

5  & DG trip & fixed trip instant(s) & $P_{DGk}\!\rightarrow\!0$ at trip time, transient in $f_{\text{mean}}(t)$, redistribution in other DGs \\

6  & \makecell[l]{Tie-line trip \\ (grid$\rightarrow$island)} & fixed opening instant & change in PCC current and frequency regulation behavior after disconnection \\

7  & \makecell[l]{Reactive power step \\ (Q-step)} & $\approx 0.50$ s & step in $Q_{\text{total}}(t)$ with voltage response and gradual $P_{\text{total}}(t)$ change \\

8  & \makecell[l]{Single-line-to-ground \\ (SLG) fault} & $\approx 0.50$ s & rise in voltage unbalance and nonzero $I_0=(I_a+I_b+I_c)/3$ \\

9  & Noise injection & full run or windowed & increased high-frequency variance in PCC voltage/current (raw vs smoothed) \\

10 & Communication delay & $\approx 0.50$ s & subtle post-event deviations in frequency and power sharing without sharp electrical transient \\

\bottomrule
\end{tabular}%
}
\end{table}

Table~\ref{tab:scenario_labels} shows all the scenarios, their label numbers, and the fixed event times used in the simulations. It also lists the main electrical signals used to verify each event. Changes in $f_{\text{mean}}(t)$, $V_{\text{PCC}}(t)$, $P_{\text{total}}(t)$, voltage unbalance, and related signals confirm that each labeled disturbance produces a clear and repeatable system response.

\subsection{Labeling Methodology and Timing}

Each scenario is assigned a unique integer label generated directly inside the digital twin and exported alongside all measurements. The label is aligned with the known event start time, ensuring that the disturbance matches the observed system response.

Let $t_n$ denote the time stamp of sample $n$, and let $\ell[n]$ denote the scenario label at sample $n$. For a disturbance that starts at time $t_e$, the label is defined as
\begin{equation}
\ell[n] =
\begin{cases}
0, & t_n < t_e \\
c, & t_n \ge t_e
\end{cases}
\label{eq:label_step}
\end{equation}
where $c \in \{1,2,\ldots\}$ is the scenario-specific class identifier. Normal operation uses $c=0$ for the full window.

For noise injection, the noisy measurement $x_{\text{noisy}}(t)$ is modeled as
\begin{equation}
x_{\text{noisy}}(t)=x(t)+\eta(t),
\label{eq:noise_model}
\end{equation}
where $\eta(t)$ is a zero-mean stochastic process.

For communication delay, the delayed feedback signal $f_{\text{fb}}(t)$ is modeled as
\begin{equation}
f_{\text{fb}}(t)=f_{\text{meas}}(t-\tau),
\label{eq:delay_model}
\end{equation}
where $\tau$ is the inserted delay.

Previously defined aggregate signals, including $P_{\text{total}}(t)$ in \eqref{eq:ptotal_qtotal} and $f_{\text{mean}}(t)$ in \eqref{eq:fmean}, are used during validation to confirm that labeled events correspond to observable system-level dynamics.

Scenario timing is deterministic across runs to ensure repeatability. This deterministic design allows direct comparison across scenarios and supports controlled benchmarking of surrogate models under both electrical disturbances and cyber-physical measurement effects.

\section{Data Export, Cleaning, and Dataset Structure}

This section explains how data from the microgrid digital twin is exported and organized into a consistent format. It also explains how the data are processed to keep them numerically stable while maintaining realistic system behavior.

\begin{figure}[!t]
    \centering
    \includegraphics[width=\linewidth]{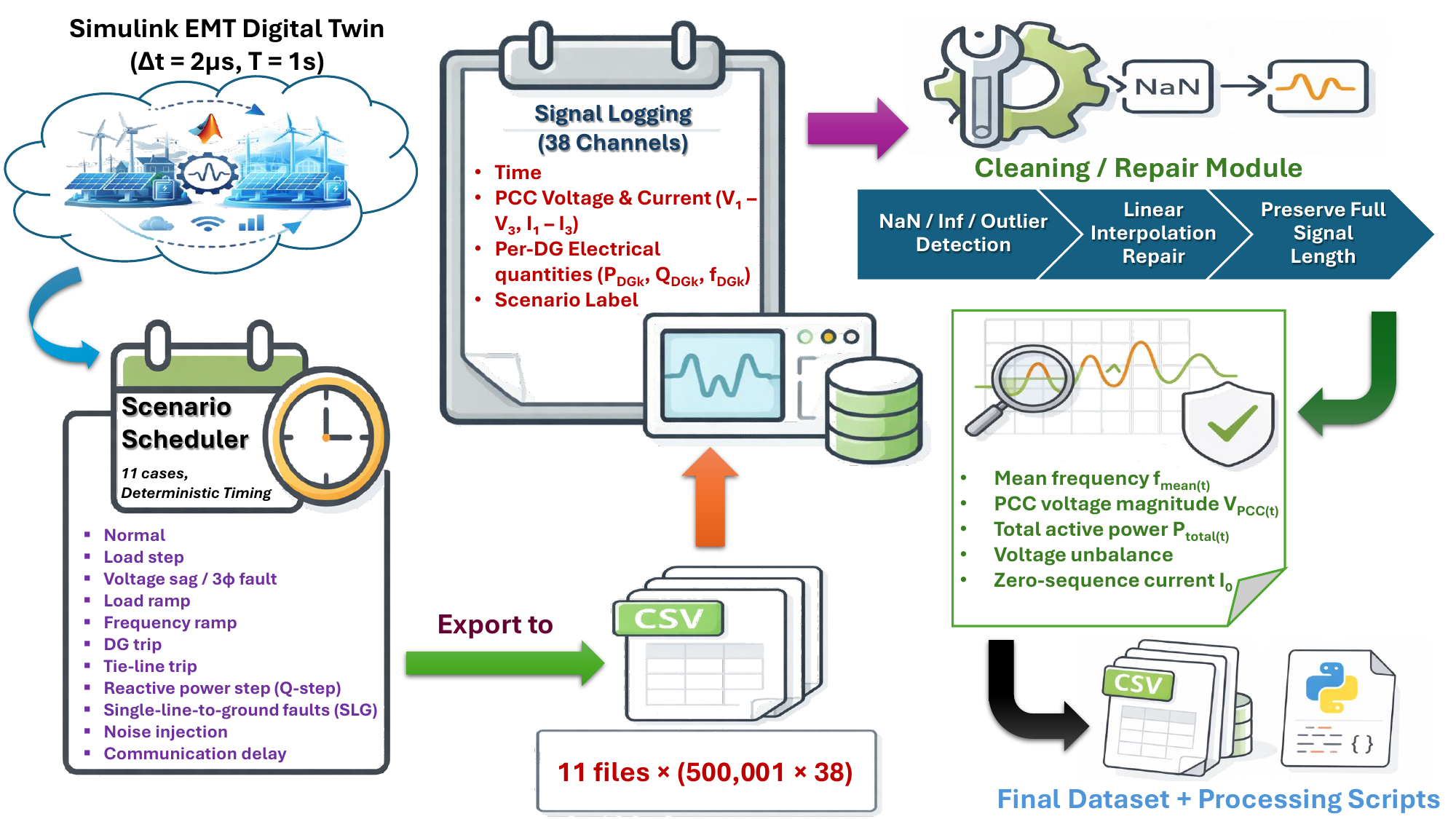}
    \caption{Dataset generation and validation pipeline. EMT simulations are performed in Simulink with fixed disturbance timing. Synchronized measurements are logged, exported to CSV, cleaned for numerical stability, and validated using system-level signals before release for surrogate modeling.}
    \label{fig:Schematic2}
\end{figure}

Figure~\ref{fig:Schematic2} shows the dataset generation and validation pipeline. The process starts with the EMT-based Simulink digital twin running at a fixed time step of $\Delta t = 2~\mu$s, consistent with high-fidelity cyber-physical energy system simulation practices \cite{palensky2017cosimulation, zhang2009power}. A disturbance scheduler applies eleven predefined operating and disturbance cases with the same timing in every run. Electrical measurements are logged in a synchronized way, including time stamps, PCC voltages and currents $(V_1,V_2,V_3,I_1,I_2,I_3)$, per-DG active and reactive power $(P_{DGk}, Q_{DGk})$, frequency $(f_{DGk})$, and scenario labels, giving a total of 38 channels per simulation.

The recorded signals are exported to CSV files and then cleaned using a repair-based method that detects and fixes invalid samples (NaN, Inf, or large outliers) with linear interpolation while keeping the full signal length. Finally, system-level checks using mean frequency $f_{\text{mean}}(t)$, PCC voltage magnitude $V_{\text{PCC}}(t)$, total active power $P_{\text{total}}(t)$, voltage unbalance, and zero-sequence current $I_0$ are used to confirm physical consistency before releasing the dataset for surrogate modeling and benchmarking.

\subsection{Signal Export and Synchronization}

All electrical and control signals are recorded directly from the microgrid digital twin EMT simulation using Simulink logging tools. Signals are saved at the original simulation time step of $\Delta t = 2~\mu$s, with no down-sampling, so fast inverter dynamics and electromagnetic transients are preserved \cite{palensky2013simulating}.

Each simulation runs for a total duration $T = 1$~s. The total number of discrete samples is therefore

\begin{equation}
N = \frac{T}{\Delta t} + 1 = 500{,}001,
\label{eq:sample_count}
\end{equation}

where the additional sample accounts for inclusion of both initial and final time instants.

The exported discrete-time signals follow

\begin{equation}
x[n] = x(n\Delta t), \quad n = 0,1,\ldots,N-1.
\label{eq:discrete_signal}
\end{equation}

For each scenario, the following synchronized signals are exported:

\begin{itemize}
\item Time stamps
\item PCC three-phase voltages $(V_1,V_2,V_3)$
\item PCC three-phase currents $(I_1,I_2,I_3)$
\item Per-DG active power $(P_{DG1} \ldots P_{DG10})$
\item Per-DG reactive power $(Q_{DG1} \ldots Q_{DG10})$
\item Per-DG frequency $(f_{DG1} \ldots f_{DG10})$
\item Scenario label
\end{itemize}

All signals use the same time vector to ensure exact alignment across all channels. The extracted signals are saved as comma-separated value (CSV) files, enabling direct use in MATLAB, Python, and machine-learning workflows.

\subsection{Dataset Schema Consistency}

All scenarios use the same dataset structure. Each CSV file contains 38 columns:

\begin{itemize}
\item 1 time column
\item 6 PCC electrical channels (three voltages and three currents)
\item 30 DG channels (10 active power, 10 reactive power, 10 frequency)
\item 1 label column
\end{itemize}

Formally, each scenario dataset is represented as a multivariate discrete-time matrix.

\begin{equation}
\mathbf{X} \in \mathbb{R}^{N \times d},
\label{eq:dataset_matrix}
\end{equation}

where $N = 500{,}001$ is the number of time samples and $d = 38$ is the number of synchronized channels. The final column corresponds to the scenario label vector

\begin{equation}
\mathbf{y} = [y[0], y[1], \ldots, y[N-1]]^{\top},
\label{eq:label_vector}
\end{equation}

with

\begin{equation}
y[n] \in \{0,1,\ldots,10\}.
\label{eq:label_set}
\end{equation}

Normal operation is assigned label 0, and each disturbance scenario is assigned a unique nonzero integer. Labels are generated directly inside the digital twin and exported together with the electrical signals to guarantee exact temporal alignment.

Using a fixed column layout keeps the input size consistent, which is required by supervised learning models.

\subsection{Data Cleaning and Repair Strategy}

The raw EMT simulation outputs can sometimes contain numerical issues, such as Not a Number (NaN) values, infinite (inf) values, or sudden spikes during switching and fault events. Instead of removing samples, a repair-based approach is adopted to preserve signal length and temporal consistency.

Let $x[n]$ denote a signal channel. Invalid samples are identified as

\begin{equation}
\mathcal{I} = \{ n \mid x[n] \notin \mathcal{D} \},
\label{eq:invalid_index}
\end{equation}

where $\mathcal{D}$ represents the predefined physically admissible range for that signal.

For each $n \in \mathcal{I}$, the repaired signal $\tilde{x}[n]$ is obtained by linear interpolation between the nearest valid samples.

\begin{equation}
\tilde{x}[n] = \text{Interp}\big(x[n_1], x[n_2]\big),
\label{eq:interpolation}
\end{equation}

where $n_1 < n < n_2$ are the closest valid indices before and after $n$. At signal boundaries, nearest-neighbor extension is applied.

For PCC voltage and current signals, out-of-range values and NaN/Inf samples are repaired using this interpolation strategy to maintain waveform continuity while preserving short transient features. All other channels are similarly checked for:

\begin{itemize}
\item NaN and Inf values
\item Invalid placeholder values
\item Extreme outliers beyond predefined physical thresholds
\end{itemize}

No rows are removed during data cleaning. Each scenario retains all $N$ samples, which preserves exact time alignment across every channel.

This repair-based preprocessing preserves underlying system dynamics while producing numerically stable inputs for surrogate model training, consistent with recommended practices for high-resolution power-system datasets \cite{aghazadeh2024digital}.

\subsection{Final Dataset Characteristics}

Each scenario produces one CSV file containing:

\begin{itemize}
\item $N = 500{,}001$ time-aligned samples
\item $d = 38$ synchronized signal channels
\item Deterministic scenario labels
\end{itemize}

Across all eleven scenarios, the dataset contains more than $5.5 \times 10^6$ labeled samples at microsecond resolution.

Unlike many public power-system datasets that focus on steady-state or low-rate measurements, this dataset preserves full electromagnetic transient behavior, inverter control dynamics, and multi-DG interactions. It is therefore suitable for training and benchmarking surrogate models for fast prediction, disturbance classification, and cyber-physical resilience analysis.

\subsection{Reproducibility Considerations}

The timing of scenarios, controller settings, and disturbance triggers is kept the same in all simulations so that the dataset can be regenerated consistently. All DG units use identical controller parameters, and events are introduced at fixed, predetermined times.

By keeping operating conditions identical, this setup enables fair comparison of surrogate models and consistent evaluation of accuracy, robustness, and generalization\cite{aravena2025open}.

\section{Scenario Validation and System-Level Evidence}

After dataset generation and cleaning, each disturbance scenario is checked using system-level electrical signals to confirm that the labels match real physical behavior. Instead of relying only on embedded labels, validation is based on frequency, PCC voltage, active and reactive power, and current measurements.

For each scenario, the disturbance timing, magnitude, and recovery are compared with expected inverter and microgrid responses. This confirms that the exported data preserves realistic control behavior and electromagnetic dynamics.

This validation step is important for data-driven modeling, since surrogate models can only learn meaningful system behavior when the signals reflect true physical responses.

\subsection{Load Step Validation}

\begin{figure}[ht]
    \centering
    \includegraphics[width=\linewidth]{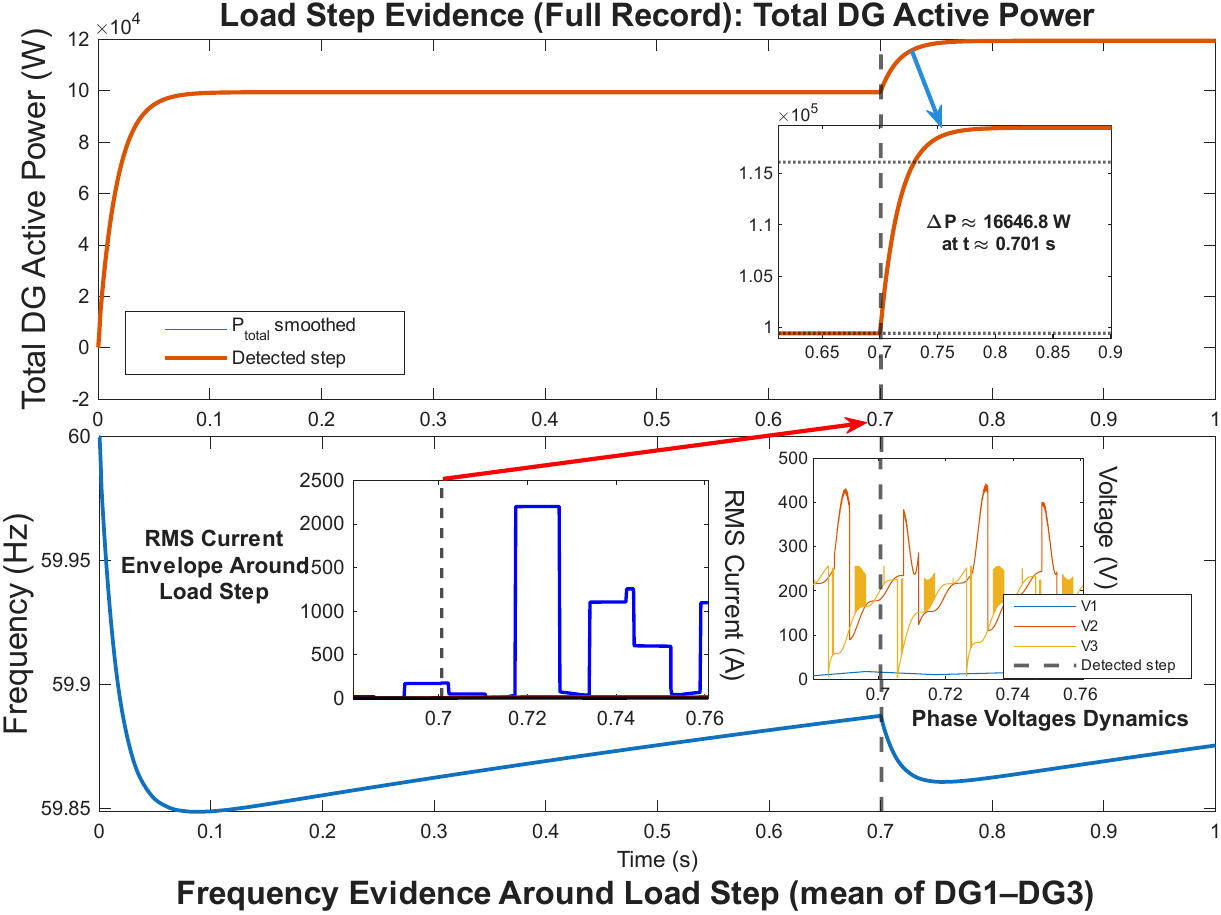}
    \caption{Load-step scenario validation and physical observability.
    The top panel shows total DG active power with an inset zoom highlighting the detected load step at approximately 0.7~s.
    The lower panel shows the mean DG frequency response, indicating an immediate dip and recovery consistent with droop control.
    Insets provide RMS current envelope and PCC phase-voltage waveforms around the step, confirming coordinated electromagnetic and control-layer responses.
   }
    \label{fig:loadstep_grouped}
\end{figure}

Figure~\ref{fig:loadstep_grouped} shows the system response to the load-step disturbance applied at about $t=0.7$~s. The top panel shows total DG active power for the whole simulation, with a zoomed view of the load step. The active power increases by about 16.6~kW and then settles to a new steady level, confirming that the load step is applied correctly.

The lower panel shows the average DG frequency, which drops right after the load step and then slowly recovers, consistent with droop control. The root-mean-square (RMS) current envelope (inset) increases at the same time, giving clear electrical evidence of the added load.

The phase voltages around the step also show short transients that line up with the power and frequency changes. Together, these signals confirm coordinated system-wide behavior. This shows that the load step is clearly captured in the dataset and produces realistic responses in power, frequency, current, and voltage, making this scenario suitable for surrogate model training.

\subsection{Voltage Sag Scenario Validation}

Three-phase PCC voltages are combined to form a voltage-magnitude proxy that represents overall electrical stress at the PCC.

Voltage sag events produce short voltage reductions followed by recovery, reflecting inverter ride-through behavior during upstream disturbances.

\begin{figure}[ht]
    \centering
    \includegraphics[width=\linewidth]{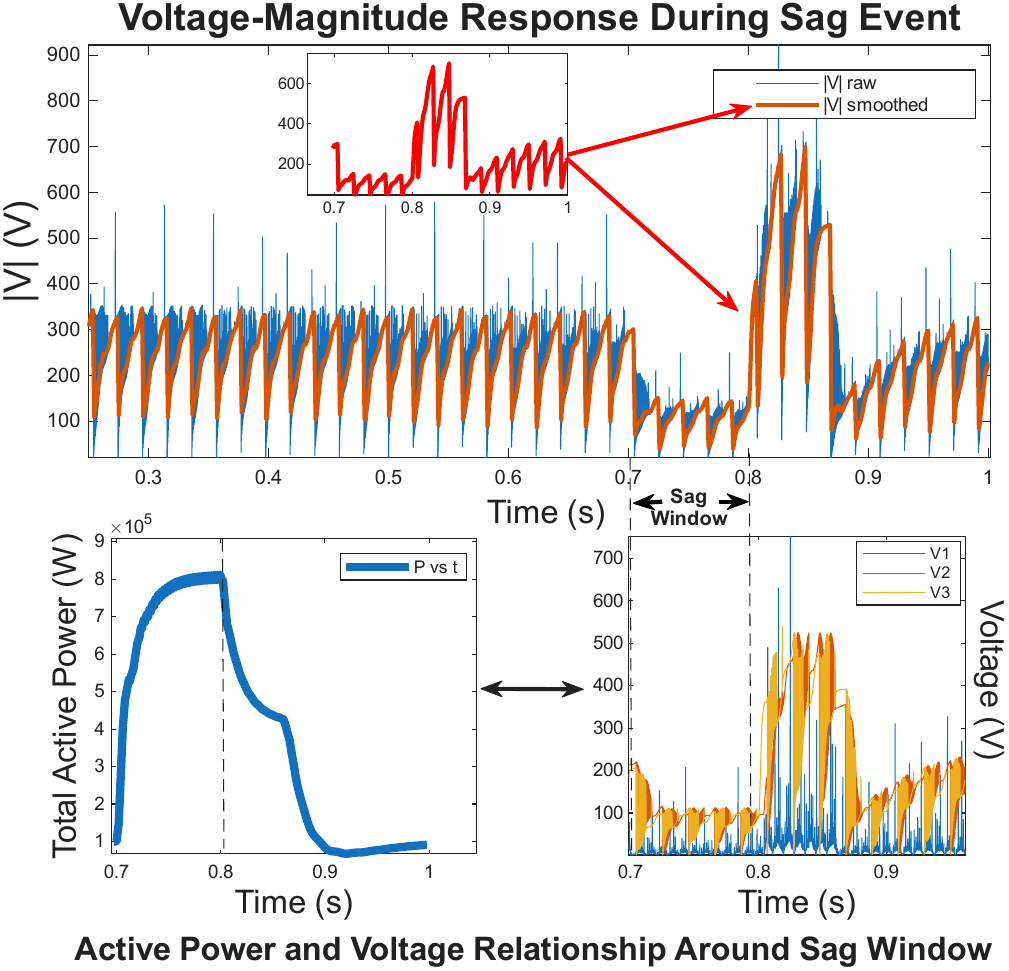}
    \caption{Voltage-sag scenario validation and physical observability.
    The top panel shows the PCC voltage-magnitude proxy (raw and smoothed) with an inset zoom highlighting the detected sag window, voltage drop, and recovery behavior.
    The lower-left panel presents the corresponding total DG active power trajectory.
    The lower-right panel shows the three-phase PCC voltages within the sag window.
    The temporal alignment across voltage and power signals confirms that the disturbance produces a coordinated system-level dynamic response.}
    \label{fig:voltagesag_grouped}
\end{figure}

Figure~\ref{fig:voltagesag_grouped} shows that the voltage sag is clearly visible in the PCC voltage magnitude, with a clear drop during the labeled sag window. Total DG active power changes during the same period, showing a brief increase followed by a noticeable reduction and gradual recovery. This behavior reflects how the inverter controllers respond to the voltage disturbance. The matching timing between voltage and total power confirms that the event affects the whole system and is correctly captured in the exported dataset.

\subsection{Load Ramp Scenario Validation}

\begin{figure}[ht]
    \centering
    \includegraphics[width=\linewidth]{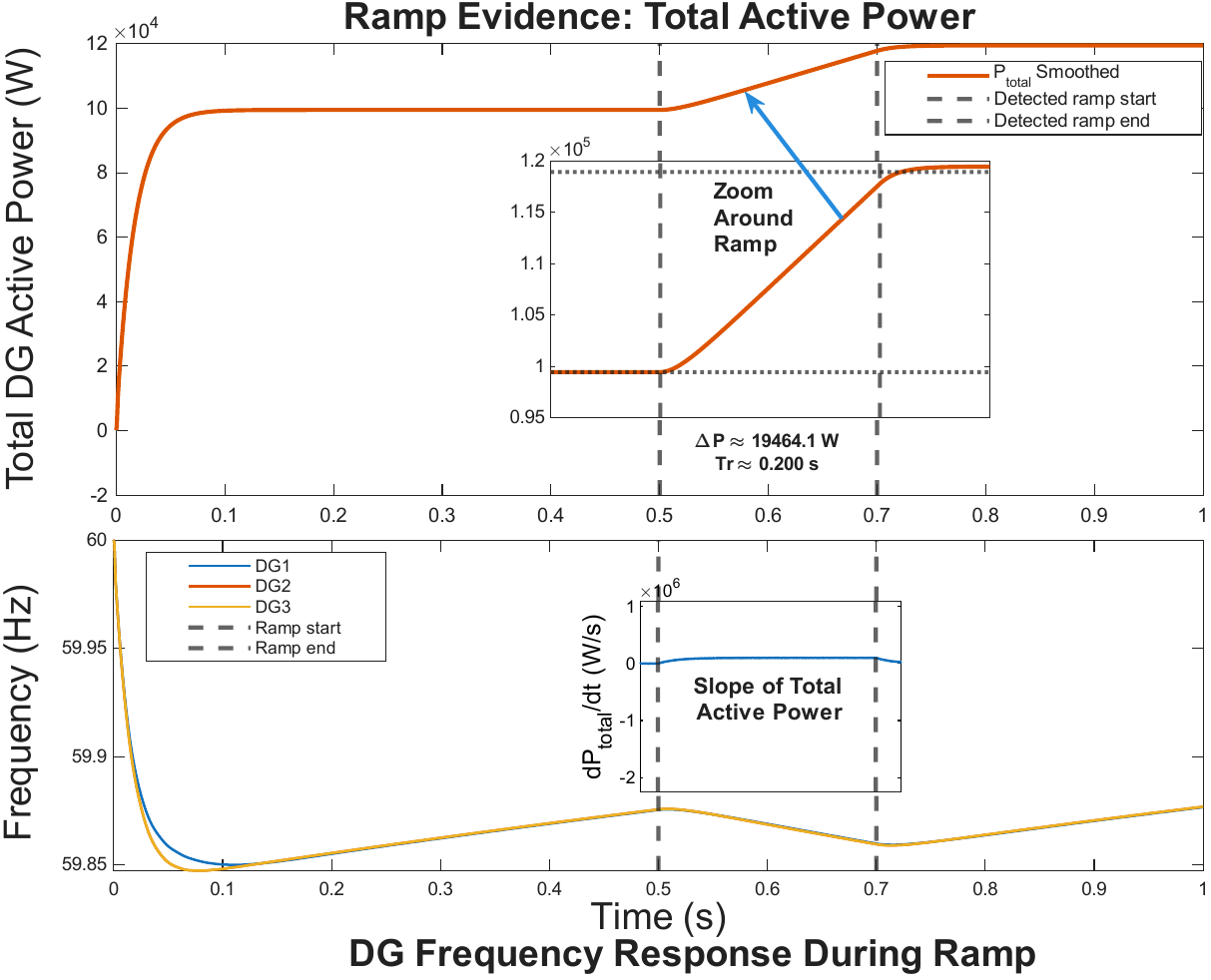}
    \caption{Load-ramp scenario validation and system-level observability.
    The top panel shows the total DG active power ($P_{\text{total}}$) in raw and smoothed form. The dashed vertical lines mark the ramp start ($t\!=\!0.5$~s) and ramp end ($t\!=\!0.7$~s). The inset zoom highlights the ramp interval and reports the estimated ramp magnitude $\Delta P$ and duration $T_r$.
    The bottom panel shows the frequency response of representative DG units during the same time window. The inset shows the slope $dP_{\text{total}}/dt$, confirming a sustained positive power ramp between 0.5--0.7~s.}
    \label{fig:loadramp_grouped}
\end{figure}

Figure~\ref{fig:loadramp_grouped} confirms that the load-ramp event is clearly visible in the exported dataset.
In the top panel, $P_{\text{total}}$ increases smoothly during 0.5--0.7~s, and then settles to a new steady level after the ramp ends.
In the bottom panel, the DG frequencies show a small but coordinated deviation during the ramp interval, which is consistent with droop-based regulation responding to a gradual change in active-power demand.
Together, the aligned changes in total active power and frequency demonstrate that the ramp produces a physically consistent system-level response suitable for surrogate model training.

\subsection{Frequency Ramp Scenario Validation}

\begin{figure}[ht]
    \centering
    \includegraphics[width=\linewidth]{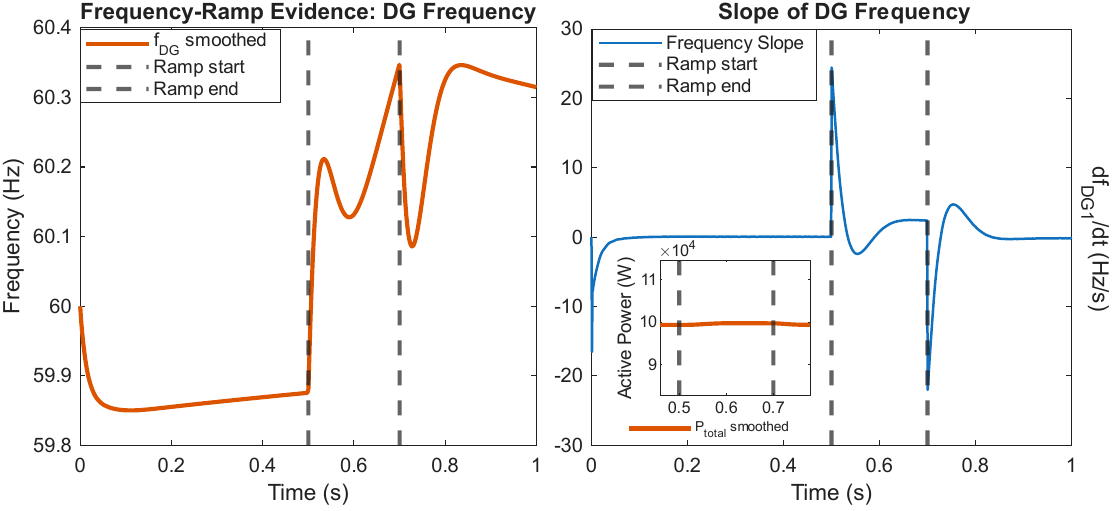}
    \caption{Frequency-ramp scenario validation and physical observability.
    Left: smoothed system frequency (DG1) showing a clear ramp between $t=0.50$~s and $t=0.70$~s, followed by transient overshoot and settling.
    Right: slope of frequency ($df/dt$), highlighting the ramp window used for labeling.
    The inset shows total DG active power remaining nearly constant during the ramp, confirming that the disturbance is frequency-driven.
    The aligned timing across frequency and slope signals demonstrates a coordinated system-level response and confirms correct realization of the frequency-ramp scenario.}
    \label{fig:freqramp_grouped}
\end{figure}

Figure~\ref{fig:freqramp_grouped} shows that the frequency-ramp disturbance is correctly applied and captured in the dataset. The left panel displays the smoothed system frequency (DG1). A clear ramp starts around $t=0.50$~s and ends near $t=0.70$~s, followed by a brief overshoot and settling to a new steady value. The total frequency increase during the ramp is about 0.3~Hz.

The right panel shows the frequency slope ($df/dt$), which clearly marks the ramp interval used for labeling. A sharp positive peak appears at the start of the ramp, followed by a negative peak near the end, reflecting inverter control action during the transition. The inset confirms that total DG active power stays almost constant during the ramp, showing that this event is driven by frequency changes rather than load variation.

These results confirm that the frequency ramp is clearly visible in both frequency magnitude and slope, produces realistic inverter control dynamics, and is correctly recorded in the exported dataset. This validates the frequency-ramp scenario for surrogate model training.

\subsection{Generator Trip Scenario Validation}

\begin{figure}[ht]
    \centering
    \includegraphics[width=\linewidth]{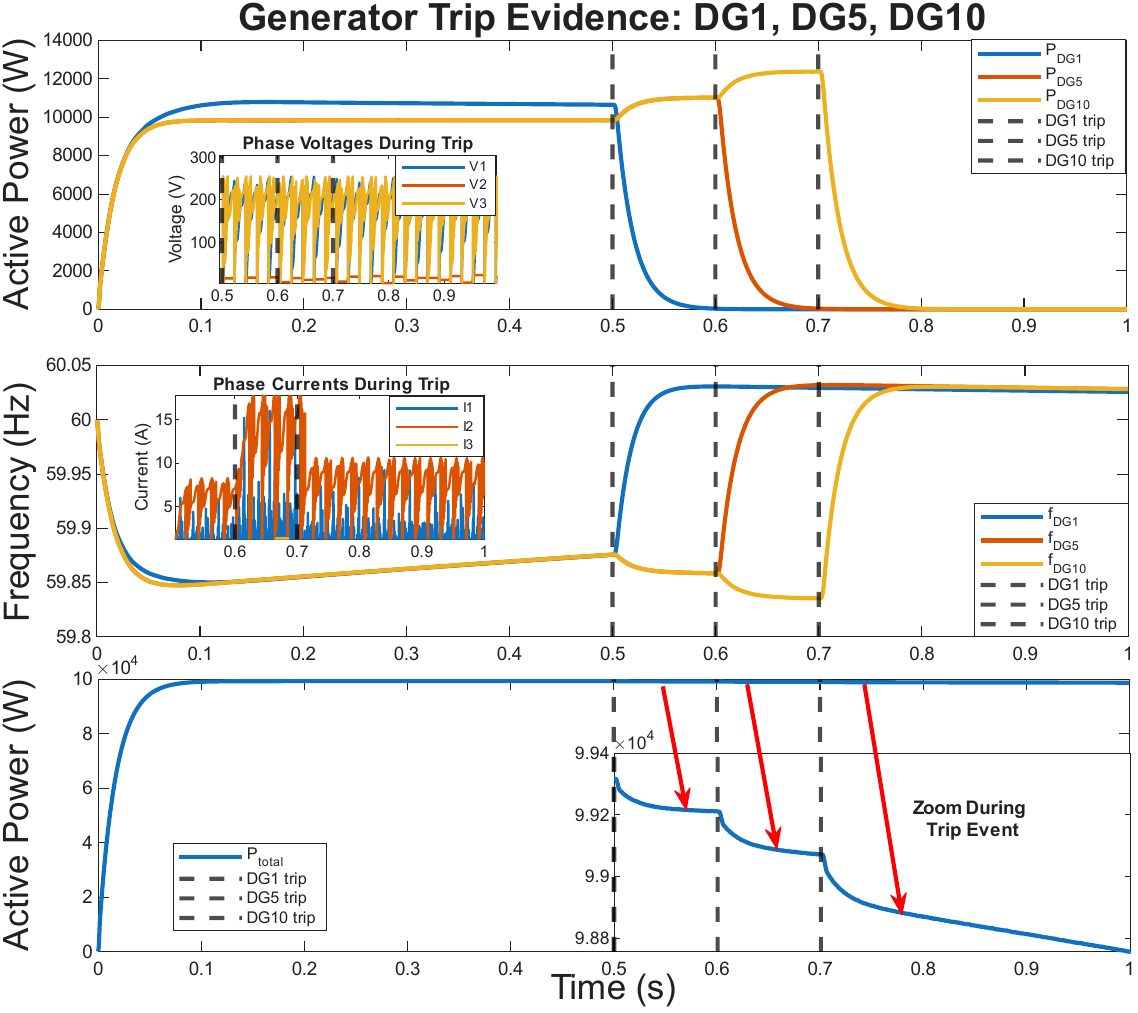}
    \caption{Generator-trip scenario validation (DG1 at $t=0.50$~s, DG5 at $t=0.60$~s, DG10 at $t=0.70$~s).
The top panel shows per-DG active power with sharp drops at each trip instant; the inset shows PCC phase-voltage transients.
The middle panel presents DG frequencies, highlighting system-wide deviation and recovery under droop regulation, with PCC phase currents shown in the inset.
The bottom panel shows total DG active power with three aligned step-like depressions corresponding to the staged outages (zoomed inset).
The synchronized responses across power, frequency, and PCC waveforms confirm correct realization and labeling of the generator-trip disturbance.}
\label{fig:gentrip_grouped}
\end{figure}

Figure~\ref{fig:gentrip_grouped} shows the staged generator-trip disturbance with three sequential outages at $t=0.50$~s (DG1), $t=0.60$~s (DG5), and $t=0.70$~s (DG10). The active power plots show that each DG quickly drops to zero at its trip time. The frequency plots show a system-wide disturbance after each trip, followed by recovery as the remaining DG units regulate the microgrid through droop control. The PCC voltage and current insets also show short electrical transients during the outages.

The total DG active power shows three clear step reductions, matching the timing of the generator trips and showing the gradual loss of generation as more units disconnect.

Other generator-trip cases, including single-DG outages across DG1–DG10 and all two-DG combinations, show the same pattern: a sudden drop in active power, followed by a frequency transient and redistribution of load among the remaining DG units. These cases were checked using the same automated detection method.

Together, these results confirm that generator-trip events are clearly visible in the dataset and are suitable for surrogate model training and disturbance classification.

\subsection{Line Trip Scenario Validation}

\begin{figure}[ht]
    \centering
    \includegraphics[width=\linewidth]{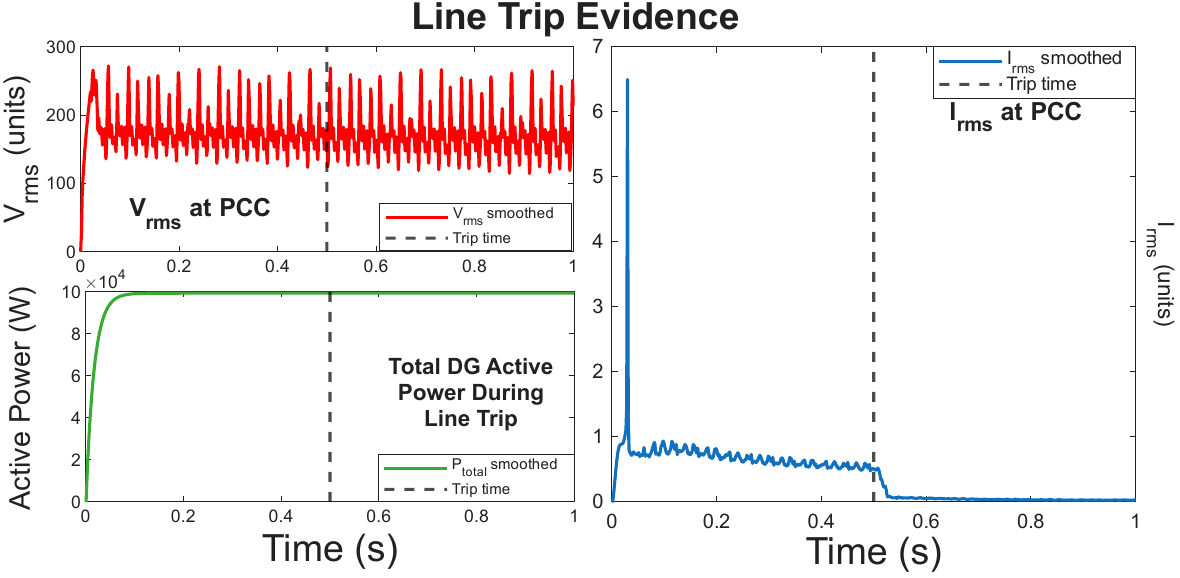}
    \caption{Line-trip scenario validation.
    The PCC RMS voltage shows a short transient at the disconnection time, while the PCC RMS current drops sharply, indicating that the line has opened.
    The total DG active power remains nearly constant before the event and displays a small depression followed by stabilization as power is redistributed.
    The aligned responses confirm correct realization and labeling of the line-trip disturbance.}
    \label{fig:linetrip_12}
\end{figure}

Figure~\ref{fig:linetrip_12} shows that the line-trip event is clearly visible in the system signals. At the disconnection time, the PCC RMS current drops sharply, showing that the line has opened, while the PCC RMS voltage shows a brief transient. At the same time, the total DG active power exhibits a small dip followed by stabilization as the microgrid redistributes power after the network separation. These aligned responses across voltage, current, and total DG power confirm coordinated system behavior and correct scenario labeling.

Other line-trip cases (DG5–DG6 and DG9–DG10) show the same pattern: a sharp change in PCC current, a brief voltage transient, and power redistribution among the remaining DG units. All cases were verified using the same automated detection metrics. Together, these results confirm that line-trip disturbances are consistently captured in the dataset and are suitable for surrogate model training and disturbance classification.

\subsection{Reactive Power Step (Q-step) Scenario Validation}

\begin{figure}[ht]
    \centering
    \includegraphics[width=\linewidth]{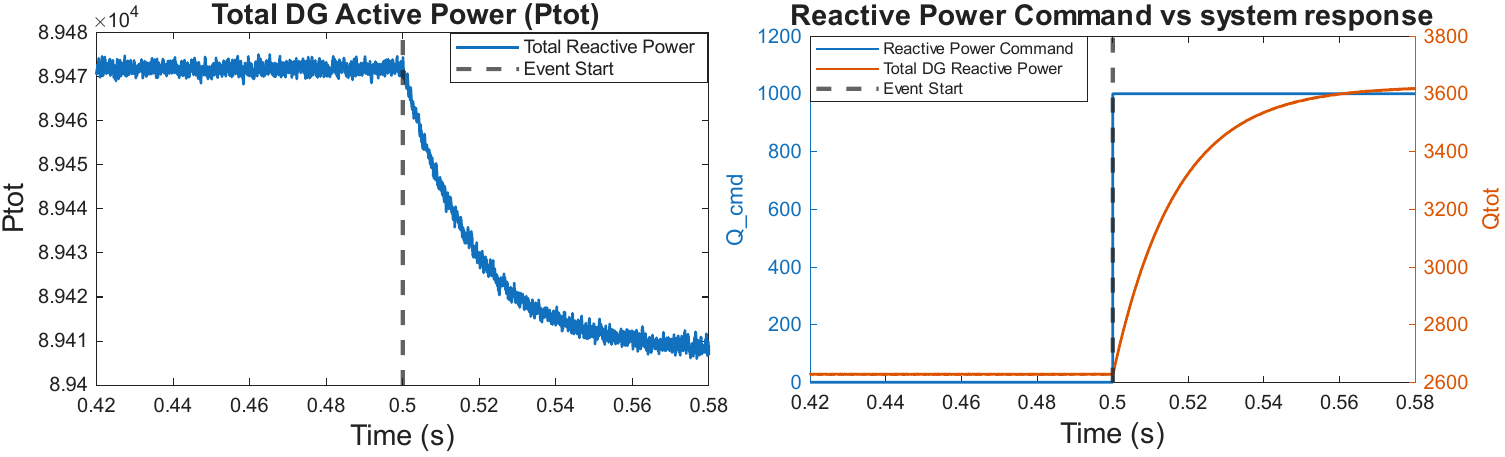}
    \caption{Reactive power step (Q-step) scenario validation. Left: total DG active power showing a gradual decrease after the event. Right: commanded reactive power step at $t = 0.5$~s and corresponding total DG reactive power response, confirming correct disturbance realization.}
    \label{fig:reactive_event_grouped}
\end{figure}

Figure~\ref{fig:reactive_event_grouped} presents the system response to a reactive power step applied at $t = 0.5$~s. Subfigure (a) shows the commanded reactive power step and the corresponding increase in total DG reactive power. The reactive power command changes abruptly at the event time, while the measured total DG reactive power rises smoothly and asymptotically toward a new steady-state value, confirming correct disturbance injection and inverter tracking behavior.

Subfigure (b) shows the total DG active power response during the same interval. Following the reactive power step, total active power gradually decreases, reflecting the inherent P--Q coupling of inverter-based resources under reactive power regulation. The active power does not exhibit a sudden discontinuity but instead transitions smoothly, consistent with controlled inverter dynamics.

Together, these results confirm that the reactive step is correctly applied at $t = 0.5$~s and produces physically consistent reactive and active power responses suitable for surrogate model training.

\subsection{Single-Line-to-Ground Fault Validation}

\begin{figure}[ht]
    \centering
    \includegraphics[width=\linewidth]{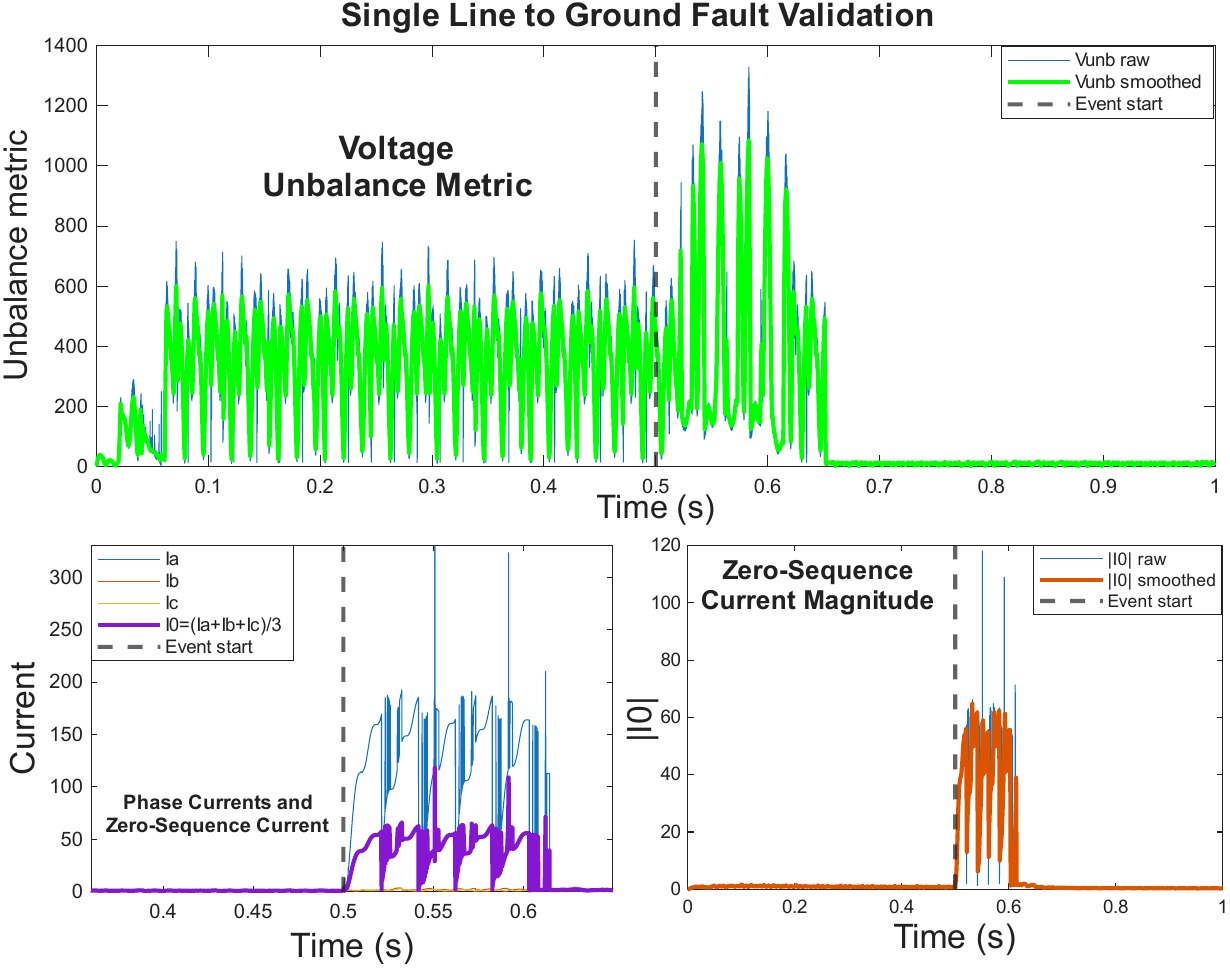}
    \caption{Single-line-to-ground (A--G) fault validation.
Top: voltage unbalance metric (raw and smoothed) with fault start at $t=0.5$~s.
Bottom-left: phase currents and zero-sequence current $I_0$.
Bottom-right: zero-sequence current magnitude $|I_0|$ (raw and smoothed).
Aligned responses confirm correct realization of the SLG disturbance.}
    \label{fig:slg_grouped}
\end{figure}

Figure~\ref{fig:slg_grouped} validates the phase A single-line-to-ground (A--G) fault, applied at $t=0.5$~s. The top panel shows the voltage unbalance metric (raw and smoothed). A clear increase begins at the fault start and continues through the fault window, then returns to near zero after clearing. This provides a compact voltage-based indicator that the system becomes strongly unbalanced during the SLG event.

The bottom-left panel presents the phase currents $(I_a, I_b, I_c)$ together with the zero-sequence current $I_0=(I_a+I_b+I_c)/3$. Before $t=0.5$~s, the currents remain close to their pre-fault values and $I_0$ stays near zero. After the fault begins, the phase currents become visibly uneven and $I_0$ rises, confirming ground involvement and the presence of zero-sequence components.

The bottom-right panel shows the zero-sequence current magnitude $|I_0|$ (raw and smoothed). A sharp increase in $|I_0|$ is observed during the same interval, followed by a return to baseline after fault clearing. This behavior matches the voltage unbalance and current responses.
Together, the rise in voltage unbalance and the appearance of zero-sequence current confirm that the A–G fault is clearly observable and correctly captured in the dataset. Similar responses are seen for B–G and C–G faults, with only the affected phase changing.

\subsection{Noise Scenario Validation}

\begin{figure}[ht]
    \centering
    \includegraphics[width=\linewidth]{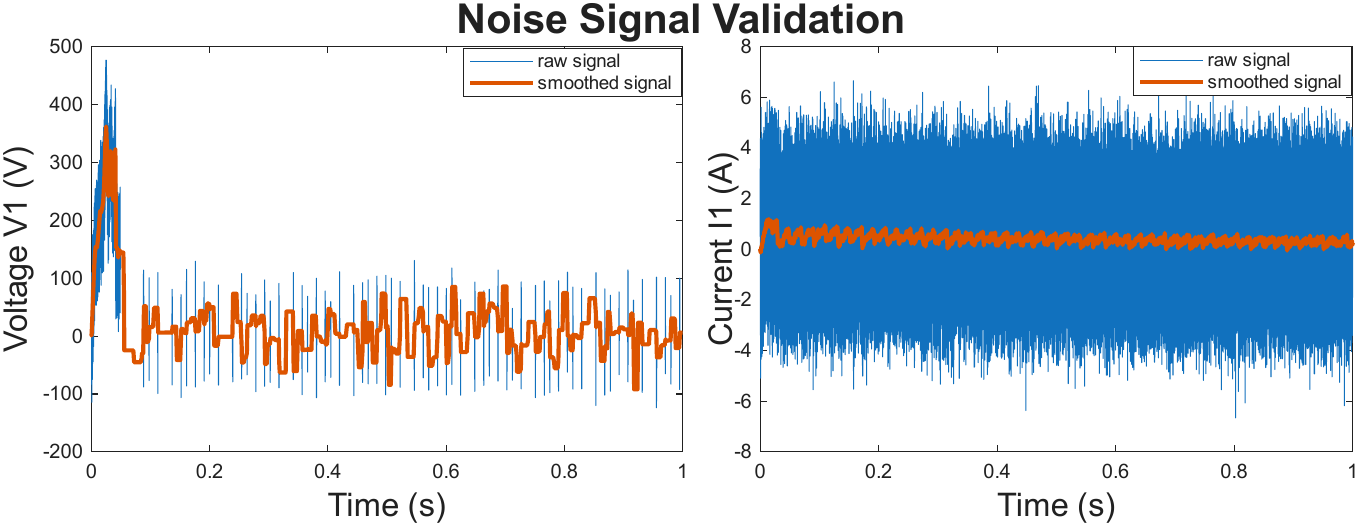}
    \caption{Noise scenario validation.
(a) PCC voltage with raw and smoothed signals under noise injection.
(b) PCC current with raw and smoothed signals.
High-frequency fluctuations in the raw measurements confirm successful noise addition, while the smoothed traces preserve the underlying system response.}
    \label{fig:noise_grouped}
\end{figure}

Figure~\ref{fig:noise_grouped} shows validation of the noise-injected dataset segment. Subfigure (a) shows the PCC voltage. The raw signal contains visible high-frequency fluctuations, while the smoothed trace keeps the main waveform. Subfigure (b) shows the corresponding PCC current, again comparing raw and smoothed signals. The raw current shows strong random variations, while the smoothed signal follows the normal trend.

These results show that noise is added to both voltage and current signals while the basic system behavior remains unchanged. By comparing the raw and smoothed signals, realistic sensor noise can be observed. This allows surrogate models to be evaluated under realistic noisy conditions.

\subsection{Communication-Delay Validation}

\begin{figure}[ht]
    \centering
    \includegraphics[width=\linewidth]{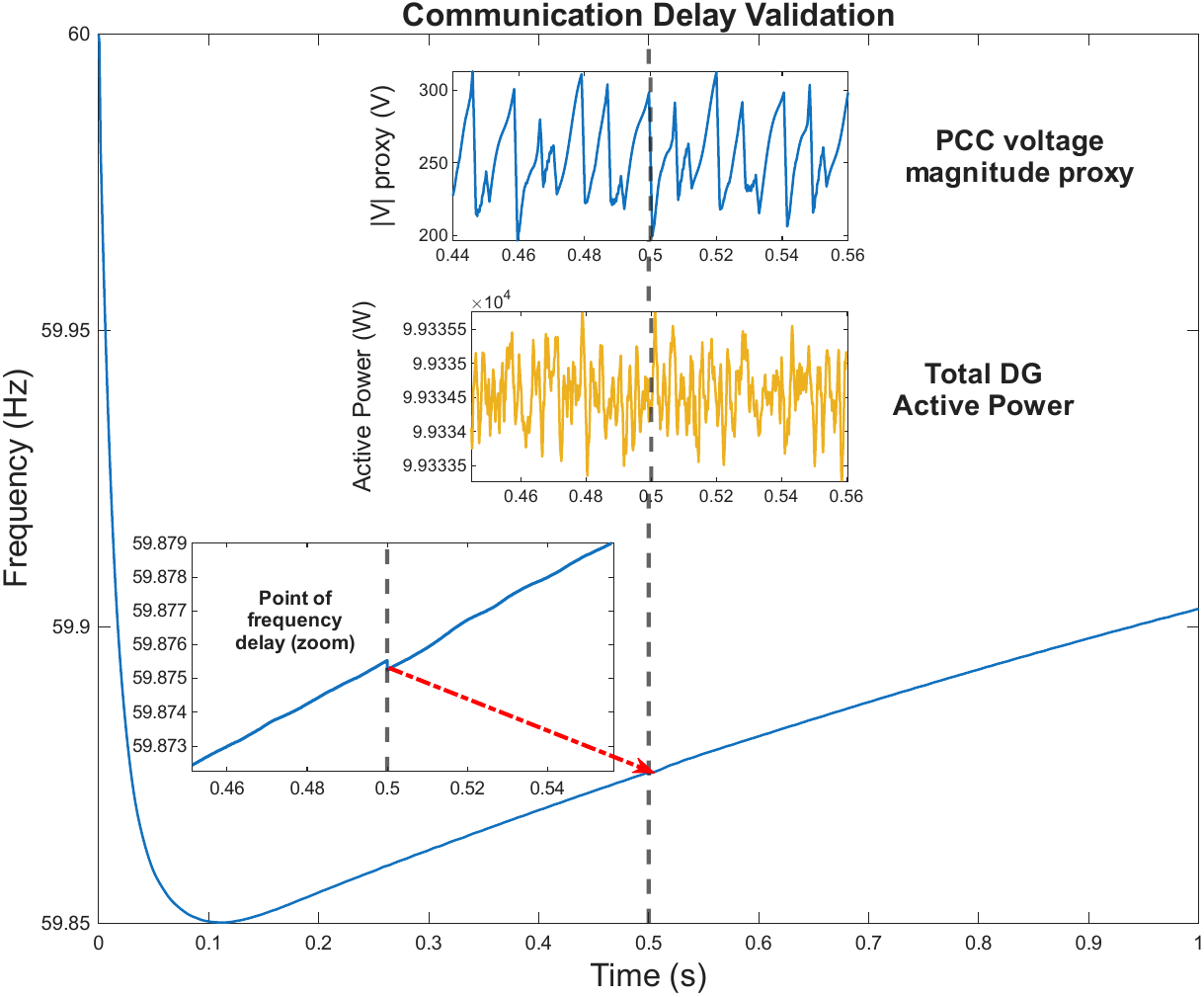}
    \caption{Communication-delay scenario validation.
Main: full system frequency response.
Lower-left inset: zoom around $t=0.5$~s showing delayed frequency response.
Upper inset: PCC voltage-magnitude proxy.
Middle inset: total DG active power.
The delay produces subtle control-related deviations rather than sharp electrical transients.}
    \label{fig:commdelay_grouped}
\end{figure}

Figure~\ref{fig:commdelay_grouped} validates the communication-delay scenario, with the delay introduced at $t=0.5$~s. The main panel shows the full system frequency response over the simulation window. Instead of a sharp transient, the frequency follows a smooth but slightly shifted trajectory after the delay is applied.

The lower-left inset zooms around $t=0.5$~s and highlights the point where delayed control action becomes visible. The frequency continues its trend but with a small change in slope, indicating the effect of delayed feedback in the control loop.

The upper inset shows the PCC voltage-magnitude proxy over the same interval. Only small variations are observed, confirming that the delay does not cause abrupt electrical disturbances. The middle inset presents the total DG active power, which shows minor fluctuations but no sudden drops or spikes.

Unlike fault or trip events, communication delay shows up as a small control effect instead of a strong electrical transient. The aligned timing across frequency, voltage, and active power confirms that the delay is correctly implemented within the control coordination layer.

\subsection{Dataset Integrity and Validation Summary}
All scenario labels are created directly inside the digital twin and exported together with the synchronized measurement signals. Automated validation scripts check the event timing and make sure the labels match the actual system behavior in every CSV file.

For all scenarios, verification is based on coordinated changes in DG frequency, PCC voltage, and total active power. These system-level signals confirm that each disturbance is physically present in the data and not just marked by a label.

By including validation in the data generation process, the dataset provides both accurate labels and realistic system dynamics. This allows the dataset to be used for surrogate modeling, fault analysis, and cyber-physical resilience studies.

\section{Discussion and Dataset Utility for Surrogate Modeling}

This dataset is designed to support surrogate modeling and cyber-physical analysis in inverter-based microgrids where fast electrical and control dynamics are critical \cite{chi2024hybrid, cedric2025quantum}. Its primary value lies not only in labeled signals, but in the fact that each label corresponds to physically observable system responses captured at electromagnetic transient resolution. Validation results throughout this paper demonstrate clear disturbance signatures in system frequency, PCC voltage, total active and reactive power, voltage unbalance, and zero-sequence current, providing measurable evidence that each scenario is correctly realized in the exported data.

Most public machine-learning datasets for power systems are sampled at low rates and mainly represent steady or slowly varying behavior \cite{ogiesoba5264265robust, islam2025machine}. While suitable for forecasting and energy management, they do not preserve inverter inner-loop dynamics, short transients, or recovery behavior during faults, islanding, or protection actions. In contrast, this dataset records synchronized measurements at a fixed EMT time step of $\Delta t = 2~\mu$s without down-sampling, capturing rapid inverter responses, brief voltage and current excursions, and frequency recovery patterns \cite{cheng2024machine}. This resolution enables learning tasks that depend on transient dynamics, including disturbance classification, short-horizon prediction, and resilience-oriented analysis.

Scenario diversity supports generalization testing. The eleven scenarios span normal operation, gradual changes (load and frequency ramps), abrupt operating transitions (DG trip and tie-line trip), balanced and unbalanced electrical disturbances (voltage sag and single-line-to-ground faults), and cyber-physical effects (noise injection and communication delay). These events produce coordinated changes in $f_{\text{mean}}(t)$, $P_{\text{total}}(t)$, $Q_{\text{total}}(t)$, $|V|_{\text{PCC}}(t)$, voltage unbalance, and zero-sequence current magnitude. As shown in the validation figures, these signals provide clear features for (i) multi-class scenario classification, (ii) regression of system-level quantities such as frequency deviation, voltage magnitude, and power redistribution, and (iii) robustness evaluation under measurement noise and delayed feedback.

All scenarios follow a consistent schema, with each CSV file containing 38 synchronized channels and a deterministic label. This fixed structure allows direct concatenation across scenarios for supervised learning without manual alignment. The dataset therefore supports waveform-based learning, sliding-window classification, and multi-output regression tasks, including prediction of $f_{\text{mean}}(t)$, $P_{\text{total}}(t)$, and $|V|_{\text{PCC}}(t)$ following disturbances \cite{lim2025power}.

Importantly, scenario validation relies on system-level observability rather than treating labels as ground truth. Each disturbance is verified using coordinated changes in frequency, PCC voltage, and power signals, which reduces the risk of training on mislabeled or physically inconsistent data. Even communication delays, introduced within the control loop, can still be observed through small but repeatable changes in frequency and power-sharing behavior. This evidence-based validation strengthens the dataset as a benchmark for surrogate models intended to learn real microgrid dynamics.

From a practical standpoint, the dataset consists of 11 CSV files, each containing 500{,}001 time-aligned samples and 38 channels, resulting in over five million labeled samples in total. While stored at microsecond resolution, typical machine-learning workflows may down-sample or window the data depending on task requirements, such as transient classification or short-term forecasting. This structure allows flexible preprocessing while preserving access to full EMT-level detail.

The dataset is also designed for reproducible benchmarking. Controller parameters, disturbance timing, and data structure are kept the same across all runs. This allows fair comparison of surrogate models, including tree-based methods, convolutional networks, recurrent models, and Transformer-based architectures. It also supports controlled robustness studies under noise and communication delay, as well as leave-one-scenario-out experiments for out-of-distribution evaluation.

A limitation is that the dataset is generated from a single microgrid topology with fixed controller settings. While this improves repeatability and interpretability, future extensions could increase diversity by varying operating points, droop gains, fault impedance and duration, grid strength, inverter limits, and communication delays. Parameter sweeps and randomized disturbances would further strengthen generalization for deployment-oriented surrogate models.

Overall, this dataset connects simplified public power-system data with realistic inverter-based microgrid behavior. By combining EMT-level measurements, multiple disturbance scenarios, deterministic labeling, and system-level validation, it provides a physically grounded benchmark for developing and testing surrogate models for transient prediction, disturbance detection, and cyber-physical resilience studies.

\section{Conclusion}

This paper presented a high-fidelity digital twin dataset for inverter-based microgrids aimed at surrogate modeling and cyber-physical analysis. The dataset is generated from an EMT microgrid model with ten inverter-based DG units and records synchronized three-phase PCC voltages and currents together with per-DG active power, reactive power, and frequency at a fixed time step of $\Delta t = 2~\mu$s. This microsecond resolution preserves fast electromagnetic transients and inverter control dynamics that are typically missing from public power-system datasets.

Eleven operating and disturbance scenarios are provided as fixed-length labeled time series, including normal operation, load step, voltage sag created by a temporary balanced three-phase fault at the PCC (cleared after a fixed duration), load ramp, frequency ramp, DG trip, tie-line trip, reactive power step, single-line-to-ground faults, measurement noise injection, and communication delay. Labels are generated inside the digital twin with deterministic timing and are validated using coordinated system-level responses in DG frequency, PCC voltage, and total DG active power, ensuring that each disturbance is physically observable in the exported data.

All scenarios share a consistent $N \times d$ structure with $N=500{,}001$ samples and $d=38$ synchronized channels, enabling direct use in machine-learning workflows. A repair-based cleaning strategy is applied in which invalid samples (NaN, Inf, or extreme outliers) are corrected using linear interpolation to preserve signal length and timing alignment. The dataset therefore supports supervised learning tasks such as disturbance classification, regression of key system variables, and waveform prediction for EMT surrogate modeling.

To support reproducible research and benchmarking, the dataset and processing scripts will be released upon acceptance. The dataset provides a basis for future evaluation of surrogate architectures, robustness under noise and delay, and runtime performance relative to the EMT digital twin. Future extensions will introduce parameter and operating-point variations to strengthen generalization and enable practical deployment of surrogate models in inverter-based microgrids.

\section*{Acknowledgments}

The authors thank the University of Tulsa Power Systems Laboratory for providing the simulation environment and computational resources used in this work.

\bibliographystyle{IEEEtran}   
\bibliography{references}      

\vfill

\end{document}